\documentclass[12pt,epsf]{article}
\usepackage{feynmf}

\newcommand{\be}{\begin{equation}}
\newcommand{\ee}{\end{equation}}
\newcommand{\ba}{\begin{eqnarray}}
\newcommand{\ea}{\end{eqnarray}}

\newcommand{\beq}{\begin{equation}}
\newcommand{\enq}{\end{equation}}
\newcommand{\bo}[1]{\mbox{\boldmath $#1$}}

\title{A perturbative approach to non-linearities in the information 
carried by a two layer neural network}

\author{{\bf\sc E. Korutcheva}
{\thanks{Regular Associate Member; Current address:
Departamento de F\'{\i}sica Fundamental,
Universidad Nacional de Educaci\'on a Distancia,
c/Senda del Rey No 9  - 28080 Madrid, Spain,
electronic address: elka@fisfun.uned.es;\,\
Permanent address: G. Nadjakov Institute of 
Solid State Physics, Bulgarian Academy of Sciences, 1784 Sofia, Bulgaria}}\\
The Abdus Salam Center for Theoretical Physics,
\\
Strada Costiera 11,
\\
34014 Trieste, Italy,
\\
\\{\bf\sc V. Del Prete} \\International School of Advanced Studies\\
Cognitive Neuroscience Sector\\ Via Beirut 2-4, 34014 Trieste, Italy
\\
\\
and \\ 
\\
{\bf\sc J.-P. Nadal} \\
Laboratoire de Physique Statistique de l'ENS\thanks{Laboratoire
associ\'e au CNRS (URA1306), \`a l'ENS et aux Universit\'es Paris 6 et Paris 7.} \\
Ecole Normale Sup\'erieure \\
24, rue Lhomond - 75231 Paris Cedex 05, France}

\date{}
\begin{document}

\maketitle

\vskip1cm
PACS: 05.20; 87.30

Keywords: Information theory, Mutual information, Infomax, Feynman diagrams

\vskip1cm
\begin{center}
Miramare, September 2000
\end{center}

\newpage

\begin{abstract}
We evaluate the mutual information  between the input and the output 
of a two layer network in the case of a noisy and non-linear analogue channel.
In the case where the non-linearity is small with respect to 
the variability in the noise, we derive an exact 
expression for the contribution to the mutual information given by the 
non-linear term in first order of perturbation theory. 
Finally we show how the calculation can be simplified by means of a 
diagrammatic expansion. Our results suggest that the use of perturbation
theories applied to neural systems might give an insight on the contribution 
of non-linearities to the information transmission and in general to the 
neuronal dynamics.   
\end{abstract}

\newpage

\section{Introduction}

The purpose of the work is to study the information properties of 
an analogue communication channel, constructed by a two-layer neural network, 
receiving data from 
a Gaussian source. This data is corrupted with 
Gaussian noise with a known variance and the output  
signals are affected by some random uncorrelated output noise.

Contrarily to what happens in the case of a linear Gaussian channel, 
which can be easily solved even in presence of 
noise \cite{vanHerten}, \cite{Guid},
the exact calculation in the case of analogue 
channel requires some assumptions on the relation between the non-linear term
and the level of noise. In particular, we suppose a small non-linearity,
compared to the output noise. This corresponds to the case where 
the sigmoidal transfer
function is relatively flat and the channel is noisy. Under this assumption, 
the mutual information between the output and the input of the channel
can be evaluated analytically. The perturbative approach by means of Feynman diagrams, 
\cite{AGD}, developed in this paper, allows to represent in a 
direct and elegant way the perturbative
corrections in first order of perturbation theory for every kind of 
non-linearity. 

Comparing with the extreme case of the binary transfer function, where
special mathematical techniques \cite{NP1}, \cite{KNP}, \cite{TKP} are
introduced for the calculation of the mutual information, 
the present analysis deals mainly
with the effect of the non-linearity on the mutual information 
and the rational way of
investigating it. The problem of its maximization with respect to the coupling
matrix \cite{Infomax} will be considered elsewhere \cite{EV}.

The paper is organized as follows: in Section 2 we introduce the model and
in Section 3 the mutual information is derived in the case 
of a general non-linear function. In Section 4 we present the results for the typical case of cubic 
non-linearities. In Section 5 we develop the rules to express the 
perturbative
series in terms of Feynman diagrams in the case of the same cubic 
non-linearity. In Section 6 we discuss the  case of a general 
non-linearity. In Section 7 we present shortly the calculation of the 
mutual information 
in the case of a generic non-local 
cubic nonlinearity and explain how the diagram technique is modified. 
We conclude with some final remarks and with future developments of this work.

\section{The network model}
We consider a two layer network with $N$ continuous inputs
$\bo x$=$\{x_1...x_N\}$ which are  Gaussian distributed and correlated
trough the matrix $C$:

\begin{eqnarray}
\langle x_i\rangle&=&0;\\
\langle x_ix_j\rangle&=&[C]_{ij}, \,\,\,\, \forall i,j\in 1,2,..N.
\label{input}
\end{eqnarray}

The signals
are corrupted by uncorrelated Gaussian input noise 
$\bo \nu$=\{$\nu_1..\nu_N\}$, with 

\begin{eqnarray}
\langle \nu_i\rangle&=&0;\\
\langle \nu_i\nu_j\rangle&=&b_0\,\delta_{ij}, \,\,\,\, \forall i,j\in 1,2,..N.
\label{inputnoise}
\end{eqnarray}

The output vector is a function of the noisy input $\bo x+\bo \nu$ transformed 
via the couplings $\{J_{ij}\}$:

\begin{equation} 
\bo V=G(J(\bo x+\bo \nu))+\bo z.
\end{equation}

We also assume that the output signals are affected by some 
random uncorrelated output noise $\bo z$=\{$z_1..z_N\}$,
with the following Gaussian distribution:

\begin{eqnarray}
\langle z_i\rangle&=&0;\\ 
\langle z_iz_j\rangle&=&b\,\delta_{ij}, \forall i,j\in 1,2,..N.
\label{outputnoise}
\end{eqnarray}

The transfer function 
$G(x)$ is a smooth continuous function, which typically  has a sigmoidal
shape in the case of analogue neuronal devices \cite{Amit}, \cite{Hertz}, \cite{Markus}.
One possible choice is:

\begin{equation}
G(x)=th(\beta x),
\label{gain}
\end{equation}

where the parameter  $\beta$ modulates the steepness of the curve.
A linear input-output relationships has already been considered in the context
of the mutual information in
previous works \cite{NP}.
Here we examine the contribution to information transmission given by a small 
non-linear term in the channel transfer function.

Assuming that the argument of the transfer function is small, a 
Taylor expansion of eq.(\ref{gain}) gives:

\begin{equation}
G(x)=th(\beta x)\simeq x-\frac{x^3}{3}+o(x^4),
\label{transfer}
\end{equation}

where the higher order terms are all odd powers of $x$.
Thus the output of the channel can be written as:

\begin{equation} 
\bo V\simeq \bo h + \bo g(\bo h)+ \bo z,
\label{output}
\end{equation}

where $\bo g(\bo h)$ is a generic non-linear term. For example it could be the 
cubic term or  a higher order term in the expansion of the 
$th(\beta x)$) in terms of
$\bo h$=$J(\bo x+\bo\nu)$.

We are interested on the mutual information \cite{Blahut} 
between the input and the output 
signals:

\begin{equation}
I=\int d\bo x\int d\bo V P(\bo x,\bo V)\log_2 \frac{P(\bo x,\bo V)}{P(\bo x)P(\bo V)}.
\end{equation}

It is easy to show that $I$ can be written as the difference between the 
output entropy and the "equivocation" between the output and the input:

\begin{equation}
I=H(\bo V)-\langle H(\bo V|\bo x)\rangle_{x},
\label{info}
\end{equation}
where

\begin{equation}
H(\bo V)=-\int d\bo V P(\bo V)\log_2 P(\bo V)
\label{entropy}
\end{equation}
and
\begin{equation}
\langle H(\bo V|\bo x)\rangle_{x}= -\int d\bo x \int d\bo V P(\bo x)P(\bo V|\bo x)\log_2 P(\bo V|\bo x).
\label{equi}
\end{equation}

In the next section we present the calculation of the mutual information
separately for
the output entropy $H(\bo V)$
and for the equivocation $\langle H(\bo V|\bo x)\rangle_x$ ,
in the considered case of a non-linear channel.

\section{The Mutual Information}
\subsection{Evaluation of the output entropy}

Let us consider the probability for the output signals $P(\bo V)$ in eq.
(\ref{entropy}). If the non-linear term $\bo g(\bo h)$, present in 
eq. (\ref{output}), were equal to zero, 
the evaluation of $H(\bo V)$ would be trivial, as $\bo V$ would be a linear
combination of Gaussian variables.
In order to extract explicitely the dependence of $P(\bo V)$ on the 
non-linear term $\bo g(\bo h)$, we introduce the conditioned 
probability $P(\bo V|\bo h)$:

\begin{eqnarray}
P(\bo V)&=&\int d\bo h P(\bo h) P(\bo V|\bo h);\\
P(\bo V|\bo h)&=&\frac{1}{\sqrt{2\pi b^N}}\cdot 
e^{-{({\bf V}-{\bf h}-{\bf g}({\bf h}))}^2/2b}.
\end{eqnarray}

Expanding $P(\bo V)$ to the first order in $\bo g$ , assuming a small 
non-linearity, compared to the variance of the output noise, we obtain:

\begin{equation}
P(\bo V)\simeq P_0(\bo V)[1+\frac{1}{b}\langle\bo 
g(\bo h)^T(\bo V-\bo h)\rangle_h + o(\bo g)],
\label{pi_0}
\end{equation}

where

\begin{eqnarray}
P_0(\bo V)&=&\int d\bo h P(\bo h) P_0(\bo V|\bo h),\\
P_0(\bo V|\bo h)&=&\frac{1}{\sqrt{2\pi b^N}}\cdot 
e^{-{({\bf V}-{\bf h})}^2/2b},\\ 
\langle F(\bo h,\bo V)\rangle_h&=
&\frac{\int d\bo h P(\bo h)P_0(\bo V|\bo h)F(\bo h,\bo V)}
{\int d\bo h P(\bo h)P_0(\bo V|\bo h)}
\end{eqnarray}

and we have assumed that higher order terms in the ratio $\bo g/b$ are 
negligible. 

Substituting eq. (\ref{pi_0}) in the expression for the output entropy 
$H(\bo V)$ we obtain 
at the first order in $\bo g$:

\begin{equation}
H(\bo V)\simeq H_0(\bo V) - \frac{1}{b}\int d\bo h\int d\bo V P(\bo h) P_0(\bo V|\bo h) \bo g(\bo h)(\bo V- \bo h) \log_2 P_0(\bo V).
\label{outfin}
\end{equation}

Here $H_0(\bo V)$ is the output entropy in the case of a linear channel \cite{vanHerten}, \cite{Guid}.

We remind that $P_0(\bo V)$ is the probability for the output $\bo V$ when $\bo g(\bo h)$=$\bo 0$. 
In this case $\bo V$ is a linear combination of zero mean Gaussian variables and its distribution is a Gaussian 
centered in $\bo 0$ with a covariance matrix given by:

\begin{equation}
\langle V_iV_j\rangle=[JCJ^T + b_0 JJ^T +b I]_{ij}=[A+bI]_{ij},
\label{Amatrix}
\end{equation}

where we have set $A=JCJ^T + b_0 JJ^T $.

$\log_2 P_0(\bo V)$\footnote{We implicitely absorb the constant \protect$1/log_e2$ in 
the definition of \protect$\log_2$ whenever we change basis from \protect$\log_2$ to \protect$\log_e$.}
can be explicitely written also in the following way:

\begin{eqnarray}
\label{I1trick}
\log_2 P_0(\bo V)&=&-\bo V^T (A+bI)^{-1}\bo V \nonumber\\
                 &=&-[ (\bo V-\bo h)^T [A+bI]^{-1} (\bo V- \bo h)+ \bo h^T [A+bI]^{-1} (\bo V-\bo h)\nonumber\\
                  &&+ (\bo V- \bo h)^T[A+bI]^{-1}\bo h^T- \bo h^T [A+bI]^{-1} \bo h,
\end{eqnarray}

which now can be easily  integrated over the Gaussian distributions $P(\bo h)$ and 
$P(\bo V|\bo h)=P(\bo V-\bo h)$. Since $\bo g(\bo h)$ is an odd power function of $\bo h$
like any term in the expansion of the transfer function (\ref{gain}), 
only the second and the third term in the sum in eq.(\ref{I1trick}) give non zero contributions.
Thus, the expression of the integral in the expression for the output entropy
eq.(\ref{outfin})
becomes:

\begin{equation}
\rightarrow\frac{1}{b}\!\!\int\!\! d\bo\ h \int\!\! d\bo V P(\bo h) P_0(\bo V|\bo h) \bo g(\bo h)^T(\bo V- \bo h)\left[\bo h^T [A+bI]^{-1} (\bo V-\bo h)+ (\bo V- \bo h)^T[A+bI]^{-1}\bo h^T\right].
\label{diagentropy}
\end{equation}

The integration over $\bo V$  leads to the final expression for the output 
entropy in terms of a general non-linearity $\bo g(h)$:

\begin{equation}
H(\bo V)\simeq H_0(\bo V) +\Delta H(\bo V),
\end{equation}

\begin{equation}
\Delta H(\bo V)=\int d\bo h P(\bo h) \bo g(\bo h)^T[A+bI]^{-1}\bo h.
\label{outentropy}
\end{equation}

The evaluation of the integral in $d\bo h$ requires a specific choice for 
the non-linearity. Before introducing it,
we  show how to obtain a similar expression for the equivocation term 
$\langle H(\bo V|\bo x)\rangle_{ x}$.

\subsection{Calculus of the equivocation term}

We remind the expression of the equivocation term:

\begin{equation}
\langle H(\bo V|\bo x)\rangle_{ x}= -\int d\bo x \int d\bo V P(\bo x)P(\bo V|\bo x)\log_2 P(\bo V|\bo x);
\end{equation}

The evaluation of this term can be carried out in a very similar way to the output entropy.

We use the equivalence:

\begin{equation}
P(\bo V|\bo x)=\int d\bo h P(\bo V|\bo h)P(\bo h|\bo x).
\end{equation}

Then, expanding $P(\bo V|\bo h)$ in powers of $\bo g(\bo h)/b$ up to the first order as 
in eq.(\ref{pi_0}) we obtain:

\begin{equation}
P(\bo V|\bo x)\simeq P_0(\bo V|\bo x)[1+\frac{1}{b}\langle\bo g(\bo h)^T(\bo V-\bo h)\rangle_{h|x} + o(\bo g)],
\label{pi_0_x}
\end{equation}

where

\begin{eqnarray}
P_0(\bo V|\bo x)&=&\int d\bo h P(\bo h|\bo x) P_0(\bo V|\bo h),\\
P_0(\bo V|\bo h)&=&\frac{1}{\sqrt{2\pi b^N}}\cdot e^{-{({\bf V}-{\bf h})}^2/2b},\\ 
\langle F(\bo h,\bo V)\rangle_{h|x}&=&\frac{\int d\bo h P(\bo h|\bo x)P_0(\bo V|\bo h)F(\bo h,\bo V)}{\int d\bo h P(\bo h|\bo x)P_0(\bo V|\bo h)}.
\end{eqnarray}

Substituting eq.(\ref{pi_0_x}) in the expression of the equivocation term, 
we obtain:

\begin{equation}
\langle H(\bo V|\bo x)\rangle_{x}\!\simeq\! 
\langle H_0(\bo V|\bo x)\rangle_{x} - \frac{1}{b}\int\!\! d\bo h\!\!\int \!\!d\bo x\!\!\int\!\! d\bo V \!P(\bo x) P(\bo h|\bo x)P_0(\bo V|\bo h) \bo g(\bo h)(\bo V\!-\! \bo h) \log_2 P_0(\bo V|\bo x).
\label{eqfin}
\end{equation}

Here the conditional probability $P_0(\bo V|\bo x)$ is:

\begin{equation}                           
\frac{1}{\sqrt{2\pi det[B+bI]}}\cdot e^{-({\bf V}- J{\bf x})^T[B+bI]^{-1}({\bf V}- J{\bf x})/2},
\end{equation} 

where $B+bI$ is the correlation matrix between the outputs in absence of signals at $\bo g=\bo 0$.
From eq. (\ref{output}),(\ref{inputnoise}),(\ref{outputnoise}) one can derive:

\begin{eqnarray} 
&&\langle (V_i- [J\bo x]_i)(V_j- [J\bo x]_j)\rangle=
[B+bI]_{ij}\nonumber\\
&&B=b_0JJ^T
\label{Bmatrix}
\end{eqnarray}

The expression for the equivocation term becomes:

\begin{equation}
\langle H(\bo V|\bo x)\rangle_{x}\simeq \langle H_0(\bo V|\bo x)\rangle_{x} +\Delta\langle H(\bo V|\bo x)\rangle_{x},
\end{equation}
where

\begin{equation}
\Delta \langle H(\bo V|\bo x)\rangle_{x}\!=\!\frac{1}{2b}\!\!\int\!\! d\bo h\!\!\int\!\! d\bo x\!\!\int\!\! d\bo V\!P(\bo x) P(\bo h|\bo x)P_0(\bo V|\bo h) \bo g(\bo h)^T(\bo V\!-\! \bo h)(\bo V\!-\! J\bo x)^T[B+bI]^{-1}(\bo V\!-\! J\bo x) .
\end{equation}

The integration over $d\bo V$ is carried easily as $P_0(\bo V|\bo h)$ is 
Gaussian and by using the replacement:

\begin{equation}
\bo V- J\bo x\rightarrow\bo V-\bo h+\bo h- J\bo x .
\end{equation}

The final expression to be integrated over $\bo V$,$\bo h$ and $\bo x$ becomes:

\begin{eqnarray}
\Delta \langle H(\bo V|\bo x)\rangle_{x}\!&=
&\!\frac{1}{2b}\!\!\int\!\! d\bo h\!\!\int\!\! 
d\bo x\!\!\int\!\! d\bo V\!\! 
P(\bo x) P(\bo h|\bo x)P_0(\bo V|\bo h) \bo g(\bo h)^T(\bo V- \bo h)
\cdot\nonumber\\
&\cdot&\left[(\bo V- \bo h)^T[B+bI]^{-1}
(\bo h- J\bo x)+(\bo h-J \bo x)^T[B+bI]^{-1}(\bo V- \bo h)\right].
\label{diagequiv} 
\end{eqnarray}

The integration over $d\bo V$ gives for the equivocation term:

\begin{equation}
\langle H(\bo V|\bo x)\rangle_{x}\simeq \langle H_0(\bo V|\bo x)\rangle_{x} + \int\! d\bo H\!\int\! d\bo x P(\bo x) P(\bo H) \bo g(\bo H+J\bo x)^T[B+bI]^{-1}\bo H ,
\label{equivoc}
\end{equation}

where we have changed variable from $\bo H\rightarrow\bo h-J\bo x$.
This expression is our final result for the equivocation term in the case of 
a general non-linear function $\bo g(h)$.

Combining eqs.(\ref{outentropy}) and (\ref{equivoc}), the mutual information 
reads:

\begin{equation}
I=I_0\!+\!\int d\bo h P(\bo h)\bo g(\bo h)^T[A+bI]^{-1}\bo h -\int\!\! d\bo H\!\!\int\!\! d\bo x P(\bo H) P(\bo x) \bo g(\bo H+J\bo x)^T[B+bI]^{-1}\bo H,
\label{finalinfo}
\end{equation}

where $I_0$ is the mutual information in absence of non-linearities
\footnote{Notice that \protect$P(\bo h)$ is different from $P(\bo H)$: both distributions are Gaussian, but with different variances,
as \protect$\bo H=\bo h-J\bo x$; \protect$\langle h_ih_j\rangle=A_{ij}$, while \protect$\langle H_iH_j\rangle=B_{ij}$. Matrices \protect$A$ and \protect$B$
are given respectively in eq.\protect(\ref{Amatrix}) and  \protect(\ref{Bmatrix}).}.

\section{Cubic non-linearity}

The final expression for the mutual information has been obtained in the 
case of a generic non-linearity
$\bo g(\bo h)$. To carry further on the calculation, we have to specify its 
shape.
Let us consider the first non-linear term in the expansion of the sigmoidal
transfer function (\ref{gain}):

\begin{equation}
\label{cub}
\bo g(\bf{h})=-g_0 \bo h^3 ,
\end{equation}

where we have set $\beta^3/3$=$g_0$.
By using the Wick theorem \cite{AGD}, and $\langle h_ih_j\rangle=A_{ij}$,
the integration over $\bf h$ in eq.(\ref{finalinfo})
can be carried out quite easily and the final expression for the output 
entropy $H(\bo V)$ for this special choice of $\bo g$ is:

\begin{equation}
\label{I1cub}
H(\bo V) =H_0(\bo V)  - 3 g_0 \sum_{i, j} A_{i i} [A+bI]^{-1}_{i j} A_{i j} .
\end{equation}

The evaluation of the integrals over $\bo x$ and $\bo H$ in 
eq.(\ref{finalinfo}) for the equivocation term can be carried out 
with the same 
procedure. As only even powers of both variables give non zero 
contribution, only the terms
 $H_{i}^{3} + 3H_{i}[J\bo x]_i[J\bo x]_i$ in the expansion
of $\bo g(\bf{H} + J\bf{x})$ remain.
The integration over $\bo x$ and $\bo H$ gives:

\begin{equation}
\label{I2cub}
\langle H(\bo V|\bo x)\rangle_{x} = \langle H_0(\bo V|\bo x)\rangle_{x} -
3g_0 \sum_{i j} [B+bI]^{-1}_{i j} [B_{i j} B_{j j} +
B_{i j} [JCJ^{T}]_{j j}] .
\end{equation}

From eqs. (\ref{I1cub}) and (\ref{I2cub}) we derive the expression for the 
mutual information:

\begin{equation}
\label{Icub}
I=I_0 - 3g_0 \sum_{i j} A_{i i}[A+bI]^{-1}_{i j}A_{i j} - [B+bI]^{-1}_{i j}B_{ij}] ,
\end{equation}

where we have used that $A-B=JCJ^{T}$ from eq.(\ref{Amatrix}), (\ref{Bmatrix}).

An interesting issue to investigate is whether the contribution to the mutual information given by higher order non-linear terms in the transfer function (\ref{gain}) is positive or negative varying the level of noise and the strength of the correlations. Here we just mention the case where the synaptic connections are positive and the inputs units are independent. In this very simple case 
it is easy to see from equation (\ref{Icub}) that the contribution to the mutual information is negative. This makes sense as the information carried by independent units is found to be additive;it is reasonable to think that a small negative non-linearity in the transfer function, which takes into account the saturation of the output to a given threshold, depresses the information. A more detailed study of the effect of a non-linearity with respect to an enhancement or depression of the mutual information will be the object of future investigations. 
 
In the limit of vanishing output noise,  $b \rightarrow 0$, by using the 

fact that $A$ and $B$ are invertible matrices, we get:

\begin{equation}
\label{Icubbsmall}
I=I_0 - 3g_0 b \sum_{i} A_{ii} [B^{-1} - A^{-1}]_{i i} + O(b^2) .
\end{equation}

\section{Diagrammatic approach for a cubic non-linearity}

It's well known from  perturbation theory \cite{AGD} that a series of Gaussian integrals can be expressed 
as a diagrammatic expansion, which makes the evaluation of high order contributions faster and elegant.
We show here how the evaluation of integrals  (\ref{diagentropy}) and (\ref{diagequiv}) can be expressed
in terms of Feynman diagrams.
Even if the formalism we develop is specific to our case, this is the first attempt to introduce a diagrammatic 
technique to take into account high order effects in information transmission in a progressive controlled way.

We summarize here the definitions and the rules which allow to build the diagrams. The general formalism can be found in \cite{AGD}.

To evaluate the output entropy in eq. (\ref{diagentropy}) we introduce the following components of the graphs and 
rules to connect them:

\begin{fmffile}{diag8}

\begin{enumerate}
\item Each term $V_i-h_i$ is represented by a wiggly line \,\,\,\,\,\ \begin{fmfgraph*}(20,1)\fmfpen{thin}\fmfleft{v1}\fmfright{o1}
\fmflabel{$i$}{v1}\fmfdot{v1}\fmf{photon,label=$V_i-h_i$}{v1,o1}\end{fmfgraph*}

\item Each term $h_i$ is represented by a solid line  \,\,\,\,\,\,\ \begin{fmfgraph*}(20,1)\fmfpen{thin}\fmfleft{v1}\fmfright{o1}
\fmflabel{$i$}{v1}\fmfdot{v1}\fmf{plain,label=$h_i$}{v1,o1}\end{fmfgraph*}

\item Each matrix element $[A+bI]_{i j}^{-1}$ is represented by a dashed square \,\,\,\,\,\,\,\,\ \begin{fmfgraph*}(20,5)
\fmfpen{thin}\fmfleft{i1}\fmfright{o1}\fmfpoly{shade,tension=0.5}{k1,k2,k3,k4}\fmflabel{$i$}{k1}\fmflabel{$j$}{k3}\fmf{phantom}{i1,k1}\fmf{phantom}{k3,o1}\end{fmfgraph*}

\item The integration over $h_i$,$h_j$ corresponds to the contraction of two solid 
lines coming out of vertices $i$,$j$, which produces the matrix element $A_{i j}$ \,\,\,\,\,\,\,\
\begin{fmfgraph*}(20,1)\fmfpen{thin}\fmfleft{v1}\fmfright{v2}\fmfdot{v1}\fmflabel{$i$}{v1}
\fmfdot{v2}\fmflabel{$j$}{v2}\fmf{plain}{v1,v2}\end{fmfgraph*}.

\item The integration over $V_i$,$V_j$ corresponds to the contraction of two wiggly 
line coming out of vertices $i$,$j$, which produces 
a term $b\delta_{i j}$ \,\,\,\,\,\,\ \begin{fmfgraph*}(20,1)\fmfpen{thin}\fmfleft{v1}\fmfright{v2}\fmfdot{v1}
\fmflabel{$i$}{v1}\fmfdot{v2}\fmflabel{$j$}{v2}\fmf{photon}{v1,v2}
\end{fmfgraph*}.
\end{enumerate}

Let us consider the case of the cubic non-linearity and let us set  
$\bo g(\bo h)=-g_0\bo h^3$.
Following the rules listed above we can identify each factor in the integrand as a diagram:

\begin{eqnarray}
\parbox{20mm}{\begin{fmfgraph*}(20,15)\fmfpen{thin}\fmfleft{v1}\fmfrightn{o}{4}
\fmfdot{v1}\fmf{plain}{v1,o1}
\fmf{plain}{v1,o2}\fmf{plain}{v1,o3}\fmf{photon}{v1,o4}\fmflabel{$i$}{v1}
\end{fmfgraph*}}\,\,\,&\longrightarrow &h_i^3(V_i - h_i)\nonumber\\
\parbox{20mm}{\begin{fmfgraph*}(20,15)\fmfpen{thin}\fmfleft{i1}\fmfright{o1}\fmfpoly{shaded,tension=0.5}{k1,k2,k3,k4}\fmf{photon,label=$i$,l.side=right}{i1,k1}\fmf{plain,label=$j$,label.side=right}{k3,o1}\end{fmfgraph*}}\,\,\,
&\longrightarrow &(V_i - h_i)[A+bI]^{-1}_{ij}h_j\nonumber\\
\parbox{20mm}{\begin{fmfgraph*}(20,15)\fmfpen{thin}\fmfleft{i1}\fmfright{o1}
\fmfpoly{shaded,tension=0.5}{k1,k2,k3,k4}\fmf{plain,label=$i$,l.side=right}{i1,k1}
\fmf{photon,label=$j$,label.side=right}{k3,o1}\end{fmfgraph*}}\,\,\,
&\longrightarrow &h_i[A+bI]^{-1}_{ij}(V_j-h_j)\nonumber
\end{eqnarray}

The result of the integrations is expressed as a series of diagrams obtained 
connecting 
the lines in the first diagram with the lines in the second and in the 
third diagram in order to construct all the topologically distinct and 
connected diagrams:

\begin{equation}
\label{diag1}
H(\bo V)\simeq H_0(\bo V)- \frac{g_{0}}{2 b} \sum_{i j k} \left<\left[\quad\parbox{20mm}
{\begin{fmfgraph*}(20,10)
\fmfleft{v1}\fmfrightn{o}{4}\fmfdot{v1}\fmf{plain}{v1,o1}\fmf{plain}{v1,o2}
\fmf{plain}{v1,o3}\fmf{photon}{v1,o4}
\fmflabel{$i$}{v1}\end{fmfgraph*}}\right]\left[\parbox{20mm}
{\begin{fmfgraph*}(20,10)\fmfpen{thin}\fmfleft{i1}
\fmfright{o1}\fmfpoly{shaded,tension=0.5}{k1,k2,k3,k4}\fmf{photon,label=$j$,label.side=right}
{i1,k1}\fmf{plain,label=$k$,label.side=right}{k3,o1}\end{fmfgraph*}}+
\parbox{20mm}{\begin{fmfgraph*}(20,10)\fmfpen{thin}\fmfleft{i1}\fmfright{o1}
\fmfpoly{shaded,tension=0.5}{k1,k2,k3,k4}\fmf{plain,label=$j$,label.side=right}
{i1,k1}\fmf{photon,label=$k$,label.side=right}{k3,o1}\end{fmfgraph*}}\right]\right>_{c} 
\end{equation}

Each of the three solid lines coming out from the first diagram can be connected with the solid line coming out from the second
diagram and similarly from the third diagram, while the remaining two solid lines are contracted in a loop;
thus we have at the end 6 times the same diagram :

\begin{equation}\parbox{40mm}
{\begin{fmfgraph*}(40,20)
\fmfpen{thin}
\fmfleft{i1}\fmfright{o1,o2}\fmf{phantom}{i1,v1}\fmf{plain,right=45}{v1,v1}
\fmfpoly{shaded,tension=0.5}{k1,k2,k3,k4}
\fmf{plain,right=0.1,tension=0.1}{v1,k1}\fmf{photon,right=0.1,tension=0.1}{k3,v1}
\fmfdot{v1}\fmf{phantom}{k1,o1}\fmf{phantom}{k3,o2}
\end{fmfgraph*}}
\end{equation}

\end{fmffile}

\begin{fmffile}{diag7}

It's  easy to check that applying the rules for the contractions of  wiggly and solid 
lines one obtains 
the expression of the output entropy which coincides with eq.(\ref{I1cub}).

Now we introduce analogous graphic rules for the evaluation of the equivocation 
(\ref{diagequiv}); some rules are the same as the ones listed, but we need a new 
element in the graph 
to represent the vector $J\bo x$.
The full prescription is given below:

\begin{enumerate}

\item Each term $V_i-h_i$ is represented by a wiggly line\,\,\,\,\,\,\,\,\,\ \begin{fmfgraph*}(20,1)\fmfpen{thin}\fmfleft{v1}\fmfright{o1}
\fmflabel{$i$}{v1}\fmfdot{v1}\fmf{photon,label=$V_i-h_i$}{v1,o1}\end{fmfgraph*}

\item Each term $H_i$ is represented by a solid line \,\,\,\,\,\,\,\,\,\,\ \begin{fmfgraph*}(20,1)\fmfpen{thin}\fmfleft{v1}\fmfright{o1}
\fmfdot{v1}\fmflabel{$i$}{v1}\fmf{plain,label=$H_i$}{v1,o1}\end{fmfgraph*}

\item Each term $[J\bo x]_i$ is represented by a dashed line \,\,\,\,\,\,\,\,\,\,\ \begin{fmfgraph*}(20,1)\fmfpen{thin}\fmfleft{v1}
\fmfright{o1}\fmfdot{v1}\fmflabel{$i$}{v1}\fmf{dots,label=$[J\bo x]_i$}{v1,o1}\end{fmfgraph*}

\item Each matrix element $[B+b]_{i j}^{-1}$ is represented by an empty square \,\,\,\,\,\,\,\,\,\,\ 
\begin{fmfgraph*}(20,5)\fmfleft{i1}\fmfright{o1}\fmfpen{thin}
\fmfpoly{empty,tension=0.5}{k1,k2,k3,k4}\fmflabel{$i$}{k1}\fmflabel{$j$}{k3}\fmf{phantom}{i1,k1}
\fmf{phantom}{k3,o1}\end{fmfgraph*}

\item The integration over $H_i$,$H_j$ corresponds to the contraction of two solid 
lines coming out of vertices $i$,$j$, which gives the matrix element $B_{i j}$  
\,\,\,\,\,\,\,\,\ \begin{fmfgraph*}(20,1)\fmfpen{thin}\fmfleft{v1}\fmfright{v2}\fmfdot{v1}\fmflabel{$i$}{v1}
\fmfdot{v2}\fmflabel{$j$}{v2}\fmf{plain}{v1,v2}\end{fmfgraph*}

\item The integration over $V_i$,$V_j$ corresponds to the contraction of two wiggly
lines coming out of vertices $i$,$j$, which gives the term $b\delta_{ij}$ 
\,\,\,\,\,\,\,\ \begin{fmfgraph*}(20,1)\fmfpen{thin}\fmfleft{v1}\fmfright{v2}\fmfdot{v1}\fmflabel{$i$}{v1}
\fmfdot{v2}\fmflabel{$j$}{v2}\fmf{photon}{v1,v2}\end{fmfgraph*}

\item The integration over $x_i x_j$ corresponds to the contraction of two dashed
lines coming out of vertices $i$,$j$, which gives the matrix element $[JCJ]^T_{ij}$\,\,\,\,\,\,\,\,\
\begin{fmfgraph*}(20,1)\fmfpen{thin}\fmfleft{v1}\fmfright{v2}\fmfdot{v1}\fmflabel{$i$}{v1}
\fmfdot{v2}\fmflabel{$j$}{v2}\fmf{dots}{v1,v2}\end{fmfgraph*}

\end{enumerate}

Let us consider the case of the cubic non-linearity. Moreover let us set
$H_i$=$h_i-[J\bo x]_i$. Then 

\begin{equation}
g_0(H_i + [J\bo x]_i)^3\rightarrow H_{i}^{3} + 3H_{i}[J\bo x]_i[J\bo x]_i
\end{equation}
as odd powers of $\bo H$ and $\bo x$ give zero contribution to the integral.

As in the case of the output entropy we can identify the different factors 
multiplied in the integrand in eq.(\ref{diagequiv}) with different diagrams:

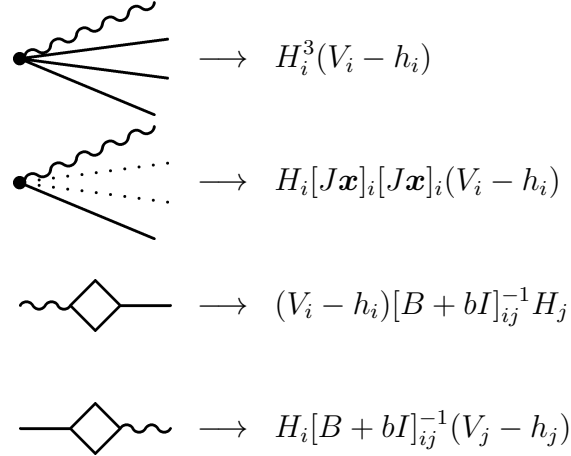
\begin{eqnarray*}
\parbox{20mm}{\begin{fmfgraph*}(20,15)\fmfpen{thin}\fmfleft{v1}\fmfrightn{o}{4}\fmfdot{v1}\fmf{plain}{v1,o1}\fmf{plain}{v1,o2}\fmf{plain}{v1,o3}\fmf{photon}{v1,o4}\fmflabel{$i$}{v1}\end{fmfgraph*}}&\longrightarrow& H_i^3(V_i - h_i)\nonumber\\
\parbox{20mm}{\begin{fmfgraph*}(20,15)\fmfpen{thin}\fmfleft{v1}\fmfrightn{o}{4}\fmfdot{v1}\fmf{plain}{v1,o1}\fmf{dots}{v1,o2}\fmf{dots}{v1,o3}\fmf{photon}{v1,o4}\fmflabel{$i$}{v1}\end{fmfgraph*}}&\longrightarrow& H_{i}[J\bo x]_i[J\bo x]_i(V_i - h_i)\nonumber\\
\parbox{20mm}{\begin{fmfgraph*}(20,15)\fmfpen{thin}\fmfleft{i1}\fmfright{o1}
\fmfpoly{empty,tension=0.5}{k1,k2,k3,k4}\fmf{photon,label=$i$,l.side=right}{i1,k1}\fmf{plain,label=$j$,l.side=right}{k3,o1}\end{fmfgraph*}}&\longrightarrow& (V_i - h_i)[B+bI]^{-1}_{ij}H_j\nonumber\\
\parbox{20mm}{\begin{fmfgraph*}(20,15)\fmfpen{thin}\fmfleft{i1}\fmfright{o1}
\fmfpoly{empty,tension=0.5}{k1,k2,k3,k4}\fmf{plain,label=$i$,l.side=right}{i1,k1}
\fmf{photon,label=$j$,l.side=right}{k3,o1}\end{fmfgraph*}}&\longrightarrow& H_i[B+bI]^{-1}_{ij}(V_j-h_j)\nonumber
\end{eqnarray*}

Thus the expression for the equivocation can be written in the following way:

\begin{eqnarray}
\label{diag2}
&&\langle H(\bo V|\bo x)\rangle_x\simeq \langle H_0(\bo V|\bo x)\rangle_x\nonumber\\
&&-\frac{g_{0}}{2 b} \sum_{i j k} \left<\left[\quad\parbox{20mm}{\begin{fmfgraph*}(20,15)
\fmfpen{thin}\fmfleft{v1}\fmfrightn{o}{4}\fmfdot{v1}\fmf{plain}{v1,o1}\fmf{plain}{v1,o2}\fmf{plain}{v1,o3}\fmf{photon}{v1,o4}\fmflabel{$i$}{v1}\end{fmfgraph*}}\quad+ 3\quad
\parbox{20mm}{\begin{fmfgraph*}(20,15)\fmfpen{thin}\fmfleft{v1}\fmfrightn{o}{4}
\fmf{plain}{v1,o1}\fmf{dots}{v1,o2}\fmf{dots}{v1,o3}\fmf{photon}{v1,o4}
\fmflabel{$i$}{v1}\end{fmfgraph*}}\right ]
\left[\parbox{20mm}{\begin{fmfgraph*}(20,15)\fmfpen{thin}\fmfleft{i1}
\fmfright{o1}\fmfpoly{empty,tension=0.5}{k1,k2,k3,k4}\fmf{photon,label=$j$,label.side=right}
{i1,k1}\fmf{plain,label=$k$,label.side=right}{k3,o1}\end{fmfgraph*}} + 
\parbox{20mm}{\begin{fmfgraph*}(20,15)\fmfpen{thin}\fmfleft{i1}\fmfright{o1}
\fmfpoly{empty,tension=0.5}{k1,k2,k3,k4}\fmf{plain,label=$j$,label.side=right}
{i1,k1}\fmf{photon,label=$k$,label.side=right}{k3,o1}\end{fmfgraph*}}\right ]\right>_{c}\nonumber . 
\end{eqnarray}

Now we have to connect both the first and the second diagram 
to the third and to the fourth diagram in all possible ways to obtain 
fully connected diagrams.

It's easy to see that the contraction of the first diagram with the third and the fourth  ones
gives $6$ times the diagram already obtained in the case of the output entropy 
(\ref{I1cub}). 

The contraction of the second diagram with the third and with the fourth diagrams 
gives a new contribution:

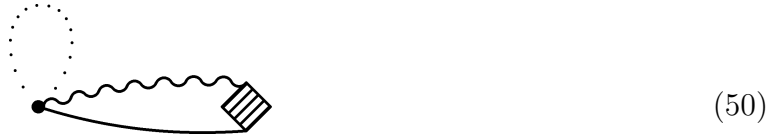
\begin{equation}\parbox{40mm}
{\begin{fmfgraph*}(40,20)
\fmfpen{thin}
\fmfleft{i1}\fmfright{o1,o2}\fmf{phantom}{i1,v1}\fmf{dots,right=45}{v1,v1}
\fmfpoly{shaded,tension=0.5}{k1,k2,k3,k4}
\fmf{plain,right=0.1,tension=0.1}{v1,k1}\fmf{photon,right=0.1,tension=0.1}{k3,v1}
\fmfdot{v1}\fmf{phantom}{k1,o1}\fmf{phantom}{k3,o2}
\end{fmfgraph*}}
\end{equation}

Writing together the two contributions we obtain the  expression for the equivocation,
which is equal to eq. (\ref{I2cub}) as it was expected.

\end{fmffile}

\section{Diagrammatic expansion and mutual information in the case of higher order 
non-linearities}

We show here how to obtain the diagrammatic expansion and the final expression for the
mutual information in the case of higher order non-linearities. This allows 
eventually to evaluate the contribution given by each term in the expansion of the 
transfer function (\ref{gain}).

Let us consider a generic term $g_{0} h_{i}^{2n+1}$ \footnote{ We always call $g_0$ the constants 
depending on parameter \protect$\beta$ not to introduce too many parameters}.

The evaluation of the integrals (\ref{diagentropy}) and (\ref{diagequiv}) can be 
carried out in a very similar way.
We make the following substitutions:

\begin{fmffile}{diag6}

\begin{eqnarray}
h_i^3&\rightarrow &h_i^{2n+1}\\
(H_i+[J\bo x]_i)^3&\rightarrow &(H_i+[J\bo x]_i)^{2n+1}
\rightarrow \sum_{l=0}^{n}\left(\begin{array}{c}2n+1\\2l\end{array}\right) (H_i)^{2n+1-2l} ([J\bo x]_i)^{2l} .
\label{rep}
\end{eqnarray}

Here the binomial expansion of $(H_i+[J\bo x]_i)^{2n+1}$ contains only even powers of $[J\bo x]_i$
because odd powers give zero contribution when integrated over $\bo x$.

These changes correspond to analogous replacements in the basic diagrams:

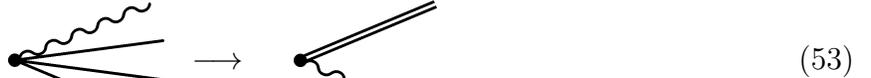
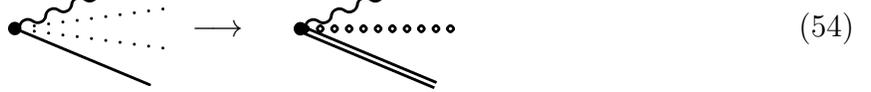
\begin{eqnarray}
\parbox{20mm}{\begin{fmfgraph*}(20,15)\fmfpen{thin}
\fmfleft{v1}\fmfrightn{o}{4}\fmfdot{v1}\fmf{plain}{v1,o1}
\fmf{plain}{v1,o2}\fmf{plain}{v1,o3}\fmf{photon}{v1,o4}
\fmflabel{$i$}{v1}\end{fmfgraph*}}
&\longrightarrow&\quad\parbox{20mm}{\begin{fmfgraph*}(20,15)\fmfpen{thin}
\fmfleft{v1}\fmfrightn{o}{2}\fmfdot{v1}\fmf{photon}{v1,o1}
\fmf{dbl_plain}{v1,o2}\fmflabel{$i$}{v1}\end{fmfgraph*}}\\  
\parbox{20mm}{\begin{fmfgraph*}(20,15)\fmfpen{thin}\fmfleft{v1}
\fmfrightn{o}{4}\fmfdot{v1}\fmf{photon}{v1,o4}\fmf{plain}{v1,o1}\fmf{dots}{v1,o2}
\fmf{dots}{v1,o3}\fmflabel{$i$}{v1}
\end{fmfgraph*}}
&\longrightarrow&\quad\parbox{20mm}{\begin{fmfgraph*}(20,15)
\fmfleft{v1}\fmfrightn{o}{3}\fmfdot{v1}\fmf{dbl_plain}{v1,o1}
\fmf{dbl_dots}{v1,o2}\fmfpen{thin}\fmf{photon}{v1,o3}\fmflabel{$i$}{v1}\end{fmfgraph*}}
\end{eqnarray}

The double solid line in the upper diagram  on the rhs is a short notation for a set of $2n+1$ 
solid lines. 

In the lower diagram on the rhs the double dotted line stands for 
a set of $2l, l=0,...,n$ single dotted lines and the double solid line 
represents a set of $2n+1-2l$ single solid lines.
Then  the diagrammatic equation for the output entropy and 
for the equivocation in the case of a generic $2n+1^{th}$ order non-linearity
can be written as follows:

\begin{equation}
\label{diagn1}
H(\bo V)\simeq H_0(\bo V)+ \frac{g_{0}}{2b} 
\sum_{i j k}\left<\left[\quad\parbox{20mm}
{\begin{fmfgraph*}(20,15)\fmfleft{v1}\fmfrightn{o}{2}\fmfdot{v1}
\fmf{dbl_plain}{v1,o1}\fmfpen{thin}\fmf{photon}{v1,o2}\fmflabel{$i$}{v1}\end{fmfgraph*}}\right]
\left[\parbox{20mm}{\begin{fmfgraph*}(20,15)\fmfpen{thin}\fmfleft{i1}\fmfright{o1}
\fmfpoly{shaded,tension=0.5}{k1,k2,k3,k4}
\fmf{photon,label=$j$,label.side=right}{i1,k1}\fmf{plain,label=$k$,label.side=right}
{k3,o1}\end{fmfgraph*}}+
\parbox{20mm}{\begin{fmfgraph*}(20,15)\fmfpen{thin}\fmfleft{i1}\fmfright{o1}
\fmfpoly{shaded,tension=0.5}{k1,k2,k3,k4}
\fmf{plain,label=$j$,label.side=right}{i1,k1}\fmf{photon,label=$k$,label.side=right}
{k3,o1}\end{fmfgraph*}}\right]\right>_{c}
\end{equation}

\begin{eqnarray}
\label{diagn2}
\langle H(\bo V|\bo x)\rangle_x&\simeq& \langle H_0(\bo V|\bo x)\rangle_x\nonumber\\
&+&\frac{g_{0}}{2b} \sum_{i j k}\left <\left [\sum_{l=0}^{n} 
\left(\begin{array}{c}2n+1\\2l\end{array}\right)\quad\parbox{20mm}{\begin{fmfgraph*}(20,15)
\fmfleft{v1}\fmfrightn{o}{3}\fmf{dbl_plain}{v1,o1}
\fmf{dbl_dots}
{v1,o2}\fmfpen{thin}\fmf{photon}{v1,o3}\fmflabel{$i$}{v1}\end{fmfgraph*}}
\right ]
\left[\parbox{20mm}{\begin{fmfgraph*}(20,15)\fmfpen{thin}\fmfleft{i1}\fmfright{o1}
\fmfpoly{empty,tension=0.5}{k1,k2,k3,k4}
\fmf{photon,label=$j$,label.side=right}{i1,k1}\fmf{plain,label=$k$,label.side=right}
{k3,o1}\end{fmfgraph*}} + 
\parbox{20mm}{\begin{fmfgraph*}(20,15)\fmfpen{thin}\fmfleft{i1}\fmfright{o1}
\fmfpoly{empty,tension=0.5}{k1,k2,k3,k4}\fmf{plain,label=$j$,label.side=right}{i1,k1}
\fmf{photon,label=$k$,label.side=right}{k3,o1}\end{fmfgraph*}}\right ]\right>_{c}\nonumber 
\end{eqnarray}

Constructing all the topologically distinct
diagrams, according to the rules given above, one can derive 
the final expression for the mutual information:

\begin{eqnarray}
\label{Igenfull}
I= &I_0&+ (2n+1)(2n-1)g_0 \sum_{i j} A_{ii}^{n}[A+b]^{-1}_{i j}A_{i j}\nonumber\\
&+&g_0 \sum_{l=0}^{2n+1} (\begin{array}{c}2n+1\\2l\end{array}) 
(2n+1-2l)\frac{1}{n}(\begin{array}{c}2n\\2\end{array})\frac{1}{l}
(\begin{array}{c}2l\\2\end{array})
\sum_{i j} [B+b]_{i j}^{-1} B_{i j} B_{i i}^{n-l}[JCJ^{T}]^{l}_{i j}.
\end{eqnarray}

Eq. (\ref{Igenfull}) is the final expression in the case of a generic non-linear
function of the type $g_i=g_0 h_{i}^{2n+1}$, for which the diagrammatic
techniques provide an easy and direct way to calculate the mutual information.
Since in the case of the sigmoidal function (\ref{gain}) the expansion includes 
only odd powers in $h_i$,  the derivation of the diagrammatic series for the whole
Taylor expansion is straightforward, at least up to the first order in $\bo g$/b.
This shows how the
diagrammatic technique provides a compact and  easily readable expression for 
the mutual information in the case of a non-linear noisy analogue channel.

\end{fmffile}

\begin{fmffile}{diag3}

\section{Generalization to  non local forms of cubic non-linearity}

Let us now investigate the case of a non local  non-linearity which depends on 
the local fields of all outputs.
 This could correspond to the
case where, for example, the global output of the network is constrained
in such a way that the local outputs of the single units depend
on the total structure of the connectivities.
The general case of $2n+1$-order non-linearities is quite complex, but the analysis can be carried out quite easily in the case of a cubic non local non-linearity. The most general third order term can be written as:

\begin{equation}
\label{gencub}
g_{i}(\bo h) = \sum_{k l m} g_{k l m}^{i} h_k h_l h_m .
\end{equation}

Substituting eq.(\ref{gencub}) in eqs.(\ref{diagentropy}) and (\ref{diagequiv})
it's easy to check that the output entropy $H(\bo V)$ and the equivocation $\langle H(\bo V|\bo x)\rangle_x$ can be written as diagrammatic equations.
The definitions for lines and vertices given in the previous section remain valid in this more complex case as well. It's enough to replace the basic diagrams derived for the cubic local non-linearity:

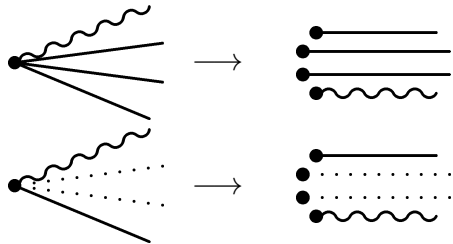
\begin{eqnarray}
\parbox{20mm}{\begin{fmfgraph*}(20,15)\fmfpen{thin}
\fmfleft{v1}\fmfrightn{o}{4}\fmfdot{v1}\fmf{plain}{v1,o1}
\fmf{plain}{v1,o2}\fmf{plain}{v1,o3}\fmf{photon}{v1,o4}
\fmflabel{$i$}{v1}\end{fmfgraph*}}
&\longrightarrow&\quad\parbox{20mm}{\begin{fmfgraph*}(20,8)\fmfpen{thin}
\fmfleft{i1,i2,i3,i4}\fmfright{o1,o2,o3,o4}\fmfdot{i1,i2,i3,i4}\fmf{photon}{i1,o1}
\fmf{plain}{i2,o2}\fmf{plain}{i3,o3}\fmf{plain}{i4,o4}\fmflabel{$i$}{i1}\fmflabel{$k$}{i2}\fmflabel{$l$}{i3}\fmflabel{$m$}{i4}
\end{fmfgraph*}}\nonumber\\  
\parbox{20mm}{\begin{fmfgraph*}(20,15)\fmfpen{thin}\fmfleft{v1}
\fmfrightn{o}{4}\fmfdot{v1}\fmf{photon}{v1,o4}\fmf{plain}{v1,o1}\fmf{dots}{v1,o2}
\fmf{dots}{v1,o3}\fmflabel{$i$}{v1}
\end{fmfgraph*}}
&\longrightarrow&\quad\parbox{20mm}{\begin{fmfgraph*}(20,8)\fmfpen{thin}
\fmfleft{i1,i2,i3,i4}\fmfright{o1,o2,o3,o4}\fmfdot{i1,i2,i3,i4}\fmf{photon}{i1,o1}
\fmf{dots}{i2,o2}\fmf{dots}{i3,o3}\fmf{plain}{i4,o4}\fmflabel{$i$}{i1}\fmflabel{$k$}{i2}\fmflabel{$l$}{i3}\fmflabel{$m$}{i4}
\end{fmfgraph*}}\nonumber
\end{eqnarray}

The diagrammatic equations for the output entropy and for the equivocation become:

\begin{equation}
\label{diag3}
H(\bo V)\simeq H_0(\bo V)+ \frac{1}{2 b} \sum_{i k l}\sum_{m \varrho\sigma} g_{k l m}^{i} \left<\left[\quad\parbox{20mm}{\begin{fmfgraph*}(20,8)\fmfpen{thin}
\fmfleft{i1,i2,i3,i4}\fmfright{o1,o2,o3,o4}\fmfdot{i1,i2,i3,i4}\fmf{photon}{i1,o1}
\fmf{plain}{i2,o2}\fmf{plain}{i3,o3}\fmf{plain}{i4,o4}\fmflabel{$i$}{i1}\fmflabel{$k$}{i2}\fmflabel{$l$}{i3}\fmflabel{$m$}{i4}
\end{fmfgraph*}}
\right]\left[\parbox{20mm}
{\begin{fmfgraph*}(20,10)\fmfpen{thin}\fmfleft{i1}
\fmfright{o1}\fmfpoly{shaded,tension=0.5}{k1,k2,k3,k4}\fmf{photon,label=$\varrho$,label.side=right}
{i1,k1}\fmf{plain,label=$\sigma$,label.side=right}{k3,o1}\end{fmfgraph*}}+
\parbox{20mm}{\begin{fmfgraph*}(20,10)\fmfpen{thin}\fmfleft{i1}\fmfright{o1}
\fmfpoly{shaded,tension=0.5}{k1,k2,k3,k4}\fmf{plain,label=$\varrho$,label.side=right}
{i1,k1}\fmf{photon,label=$\sigma$,label.side=right}{k3,o1}\end{fmfgraph*}}\right]\right>_{c} 
\end{equation}

\begin{eqnarray}
\label{diag4}
&&\langle H(\bo V|\bo x)\rangle_x\simeq \langle H_0(\bo V|\bo x)\rangle_x\nonumber\\
&&+\frac{1}{2 b} \sum_{i k l}\sum_{m \varrho \sigma} \left<\left [\quad\parbox{20mm}{\begin{fmfgraph*}(20,8)\fmfpen{thin}
\fmfleft{i1,i2,i3,i4}\fmfright{o1,o2,o3,o4}\fmfdot{i1,i2,i3,i4}\fmf{photon}{i1,o1}
\fmf{plain}{i2,o2}\fmf{plain}{i3,o3}\fmf{plain}{i4,o4}\fmflabel{$i$}{i1}\fmflabel{$k$}{i2}\fmflabel{$l$}{i3}\fmflabel{$m$}{i4}
\end{fmfgraph*}}\quad+\quad\parbox{20mm}{\begin{fmfgraph*}(20,8)\fmfpen{thin}
\fmfleft{i1,i2,i3,i4}\fmfright{o1,o2,o3,o4}\fmfdot{i1,i2,i3,i4}\fmf{photon}{i1,o1}
\fmf{plain}{i2,o2}\fmf{dots}{i3,o3}\fmf{dots}{i4,o4}\fmflabel{$i$}{i1}\fmflabel{$k$}{i2}\fmflabel{$l$}{i3}\fmflabel{$m$}{i4}
\end{fmfgraph*}}
\right ]
\left[\parbox{20mm}{\begin{fmfgraph*}(20,15)\fmfpen{thin}\fmfleft{i1}
\fmfright{o1}\fmfpoly{empty,tension=0.5}{k1,k2,k3,k4}\fmf{photon,label=$\varrho$,label.side=right}
{i1,k1}\fmf{plain,label=$\sigma$,label.side=right}{k3,o1}\end{fmfgraph*}} + 
\parbox{20mm}{\begin{fmfgraph*}(20,15)\fmfpen{thin}\fmfleft{i1}\fmfright{o1}
\fmfpoly{empty,tension=0.5}{k1,k2,k3,k4}\fmf{plain,label=$\varrho$,label.side=right}
{i1,k1}\fmf{photon,label=$\sigma$,label.side=right}{k3,o1}\end{fmfgraph*}}\right ]\right>_{c}\nonumber 
\end{eqnarray}

Following the rules for the contraction of the wiggly and solid lines it's easy to derive the final expression for the mutual information:

\begin{equation}
\label{Igencub}
I = I_0 +3 \sum_{i k l}\sum_{\varrho m} A_{k l} g_{k l m}^{i} [[A+bI]^{-1}_{m \varrho}A_{\varrho i} -
[B+bI]^{-1}_{m \varrho}B_{\varrho i}]
\end{equation}

We list some specific cases arising from this generic nonlinearity and the correspondent
final expression for the mutual information:
\\
\vskip0.5cm
{\bf case 1}
\begin{equation}
\label{c1}
g_{k l m}^{i} = g_{0} \delta_{k i} \delta_{l i} \delta_{m i},
\end{equation}
leading to the case already analyzed:$\sum_{k l m}g_{k l m}^{i}h_k h_l h_m= g_{0} h_{i}^{3}$.
The expression for the mutual information is given by eq.(\ref{Icub}).
\\
\vskip0.5cm
{\bf case 2}
\begin{equation}
\label{c2}
g_{k l m}^{i} = g_{0} \delta_{m i} \delta_{k l} \rightarrow \sum_{k l m}g_{k l m}^{i}h_k h_l h_m=
g_{0} h_{i} 
(\sum_{k} h_{k}^{2}).
\end{equation}

\begin{equation}
I = I_0 + g_{0} [Tr A Tr D + 2 Tr AD]
\end{equation}
where $D=[I+bA^{-1}]^{-1} - [I+bB^{-1}]^{-1}$.
\\
\vskip0.5cm
{\bf case 3}

\begin{equation}
\label{c3}
g_{k l m}^{i} = g_{0} \rightarrow \sum_{k l m}g_{k l m}^{i}h_k h_l h_m=g_{0} (\sum_{k} h_{k})^3
\end{equation}
 
\begin{equation}
I = I_0 + 3 g_{0} (\sum_{k l} A_{k l})(\sum_{m i} D_{m i}) 
\end{equation}
\\
\vskip0.5cm
{\bf case 4}
\begin{equation}
\label{c4}
g_{k l m}^{i} = g_{0} \delta_{k i} \delta_{l i} 
\rightarrow \sum_{k l m}g_{k l m}^{i}h_k h_l h_m=g_{0} h_{i}^{2} \sum_{m} h_m 
\end{equation}

\begin{equation}
I = I_0 + g_{0} \sum_{i} (A_{i i} \sum_{m} D_{i m} + 2 D_{i i} \sum_{m} A_{i m})
\end{equation}
\\
\vskip0.5cm
{\bf case 5}
\begin{equation}
\label{c5}
g_{k l m}^{i} = g_{0} \delta_{k l} \rightarrow \sum_{k l m}g_{k l m}^{i}h_k h_l h_m=
g_{0}(\sum_{k} h_{k}^{2})(\sum_{m} h_m)
\end{equation}

\begin{equation}
I = I_0 + g_{0}\left(TrA \sum_{i m} D_{m i} + \sum_{i m} ([AD]_{m i} + [DA]_{m i})\right).
\end{equation}
\\
\vskip0.5cm
{\bf case 6}
\begin{equation}
\label{c6}
g_{k l m}^{i} = g_{0} \delta_{k l} \delta_{l m} \rightarrow\sum_{k l m}g_{k l m}^{i}h_k h_l h_m= g_{0} \sum_{k}h_{k}^{3}
\end{equation}

\begin{equation}
I = I_0 + 3g_{0} \sum_{i k} A_{k k}D_{k i} .
\end{equation}

\section{Final remarks}

In the present paper we have developed a perturbative approach for the calculation
of the mutual information in the case of  a generic non-linear channel by means of  
Feynman diagrams.
As far as we know, this is the first attempt to use this techniques 
in the context of the mutual information.

Our analysis is valid in the case of small non-linearity compared to the
output noise and possibly for any flat sigmoidal 
transfer function of  a noisy channel.

We show  systematically how the consecutive steps to calculate the 
mutual information can be
easily performed introducing proper diagrammatic rules, in analogy to other
standard perturbative approaches \cite{AGD}.

We investigate more in detail the case of
{\em local} non-linear transfer functions, when the
output of each unit depends only on its local field. 
Previous works have shown that this regime provides an optimal information transfer \cite{NP}.
Then we apply the same techniques to the more general case of 
{\em non-local} non-linearities, restricted to cubic powers of $\bo h$,
where the output of each unit depends on the total structure of the connectivities.
This regime corresponds to the case where the total output of the network is constrained 
in such a way that the state of each output unit can be modified by any pair 
interaction.

Further developments of this analysis include the maximization of the mutual
information 
with respect to the coupling matrix $J$
in order to find the optimal structure of the connectivities. This should hopefully
provide more 
interesting results, compared to the linear case, \cite{Guid}, and it will be the object
of future investigations \cite{EV}.

\vskip0.5cm
{\bf Acknowledgments} E.K. warmly thanks for hospitality and support
the Abdus Salam International Center
for Theoretical Physics, Trieste, Italy, where this work
was completed.
V.D.P. thanks A.Treves, Stefano Panzeri, Giuseppe Mussardo and Ines Samengo for useful discussions.  
The work is also supported by the Spanish DGES Grant 
PB97-0076 and partly by Contract F608 with the
Bulgarian Scientific Foundation.


\end{fmffile}
\end{document}